\documentclass[10pt,conference]{IEEEtran}
\IEEEoverridecommandlockouts

\usepackage{cite}
\usepackage{amsmath,amssymb,amsfonts}
\usepackage{algorithmic}
\usepackage{graphicx}
\usepackage{textcomp}
\usepackage{xcolor}
\usepackage{caption}
\usepackage{subcaption}
 
\usepackage{float} 
\def\BibTeX{{\rm B\kern-.05em{\sc i\kern-.025em b}\kern-.08em
    T\kern-.1667em\lower.7ex\hbox{E}\kern-.125emX}}

\usepackage[braket]{qcircuit}
\usepackage{amsthm}

\newtheorem{definition}{Definition}

\begin{document}

\title{NovaQ: Improving Quantum Program Testing through Diversity-Guided Test Case Generation}


\author{\IEEEauthorblockN{Tiancheng Jin}
\IEEEauthorblockA{\textit{Kyushu University, Japan} \\
jintc1@f.ait.kyushu-u.ac.jp}
\and
\IEEEauthorblockN{Shangzhou Xia}
\IEEEauthorblockA{\textit{Kyushu University, Japan} \\
xia.shangzhou.218@s.kyushu-u.ac.jp}
\and
\IEEEauthorblockN{Jianjun Zhao\thanks{This work was supported by JSPS KAKENHI Grant No. JP24K02920, No. JP23K28062, No. JP24K14908, and JST SPRING Grant No. JPMJSP2136. }}
\IEEEauthorblockA{\textit{Kyushu University, Japan} \\
zhao@ait.kyushu-u.ac.jp}
}

\maketitle

\begin{abstract}
Quantum programs are designed to run on quantum computers, leveraging quantum circuits to solve problems that are intractable for classical machines. As quantum computing advances, ensuring the reliability of quantum programs has become increasingly important. This paper introduces \textit{NovaQ}, a diversity-guided testing framework for quantum programs. \textit{NovaQ} combines a distribution-based test case generator with a novelty-driven evaluation module. The generator produces diverse quantum state inputs by mutating circuit parameters, while the evaluator quantifies behavioral novelty based on internal circuit state metrics, including magnitude, phase, and entanglement. By selecting inputs that map to infrequently covered regions in the metric space, \textit{NovaQ} effectively explores under-tested program behaviors. We evaluate \textit{NovaQ} on quantum programs of varying sizes and complexities. Experimental results show that \textit{NovaQ} consistently achieves higher test input diversity and detects more bugs than existing baseline approaches.

\end{abstract}


\section{Introduction}

Quantum programs are designed to run on quantum computers, leveraging quantum circuits to solve problems that are intractable for classical machines~\cite{kitaev2002classical}. They are widely used in domains such as cryptography~\cite{gisin2002quantum}, optimization~\cite{abbas2024challenges}, and quantum simulation~\cite{georgescu2014quantum}. As quantum hardware and software evolve, the increasing complexity of quantum circuits makes ensuring the correctness and robustness of quantum programs critical~\cite{piani2016robustness, harrow2003robustness, jin2023scaffml}, particularly given the high cost of quantum computation~\cite{streltsov2012quantum} and the susceptibility to subtle errors~\cite{gottesman2002introduction}.

Despite growing interest in quantum computing, systematic testing of quantum programs remains challenging and underdeveloped~\cite{leite2025testing,miranskyy2019testing}. Traditional testing techniques are often ineffective for quantum programs due to the probabilistic and non-deterministic nature of quantum behavior~\cite{paltenghi2024survey}. Existing approaches usually rely on randomly generated inputs~\cite{ye2023quratest}, which lack diversity and make bug detection difficult. These challenges require new testing methodologies that can efficiently explore the quantum input space and identify faults.

This paper presents \textit{NovaQ}, a testing framework designed specifically for quantum programs. NovaQ consists of two core components: a distribution-based test case generator and a novelty-driven test case evaluator. The generator creates test inputs by sampling gate parameters from Gaussian distributions~\cite{zhang2016gaussian}, applying them to a parameterized initial quantum circuit, and executing the circuit in the all-zero quantum state to produce quantum state vectors. To maintain diversity, NovaQ employs a mutation-based strategy that perturbs the mean and variance of the parameter distributions throughout iterations.

To evaluate test cases, NovaQ analyzes internal circuit state metrics—such as magnitude, phase, and entanglement—to compute a novelty score that quantifies behavioral diversity. Test inputs that explore novel areas of the state space are prioritized in subsequent iterations. This feedback loop enables NovaQ to efficiently explore diverse circuit behaviors and uncover bugs in quantum programs.

We evaluate \textit{NovaQ} on a set of quantum programs with varying circuit sizes and complexities. Specifically, we apply NovaQ to extend the test case generator proposed by Ye \textit{et al.} in QuraTest~\cite{ye2023quratest}, adjusting the three parameters of the U gate to generate diverse test cases. The number of qubits in the generated test cases ranges from 3 to 12. These test cases are used as inputs to benchmark programs to assess their bug-detection capability. Experimental results show that NovaQ consistently outperforms baseline testing approaches in terms of both test case diversity and fault detection. By combining guided test case generation with diversity metrics tailored to quantum behavior, NovaQ provides a promising approach for enhancing the reliability of quantum software systems.

\section{Background}\label{sec:Bac}

This section introduces essential background concepts related to quantum programs and quantum circuits.

\subsection{Quantum Bits}

The fundamental elements in quantum programs are quantum bits, commonly referred to as \textit{qubits}. A classical bit can take the value 0 or 1, while a qubit can exist in a linear superposition of these two basis states. The basis states $\ket{0}$ and $\ket{1}$ are represented as $\ket{0} = [1, 0]^\top$ and $\ket{1} = [0, 1]^\top$, and are called computational basis states. A general qubit state is written as $\ket{q} = \alpha\ket{0} + \beta\ket{1}$, where $\alpha$ and $\beta$ are complex numbers satisfying the \textit{normalization condition} $\left| \alpha \right|^2 + \left| \beta \right|^2 = 1$. 

\subsection{Quantum Gates and Circuits}

Quantum gates manipulate the states of qubits and form the building blocks of quantum circuits. Similar to gates in classical digital circuits, quantum gates are linear and reversible, operating on a fixed number of qubits with equal numbers of input and output lines.

This paper uses the $U$ gate, a commonly used single-qubit gate, defined as:
\[
U(\theta, \phi, \lambda) =
\begin{pmatrix}
\cos\left(\frac{\theta}{2}\right) & -e^{i\lambda} \sin\left(\frac{\theta}{2}\right) \\
e^{i\phi} \sin\left(\frac{\theta}{2}\right) & e^{i(\phi + \lambda)} \cos\left(\frac{\theta}{2}\right)
\end{pmatrix}
\]

Quantum circuits represent executable quantum programs. They consist of an ordered sequence of quantum gates applied to qubits, followed by measurement operations. 
The output of a quantum circuit is obtained by measuring the final state of the qubits.

\subsection{Quantum Program Module}



Similar to classical modules that encapsulate code for specific functionality, quantum program modules encapsulate reusable units composed of quantum gate sequences, subcircuits, measurement operations, or subalgorithm implementations. For instance, the Inverse Quantum Fourier Transform (IQFT) module~\cite{weinstein2001implementation} is a widely used module for converting phase information into magnitude information.
\section{Methodology}

\subsection{Overview}

\begin{figure}[h]
    \centering
    \includegraphics[width=1.0\linewidth]{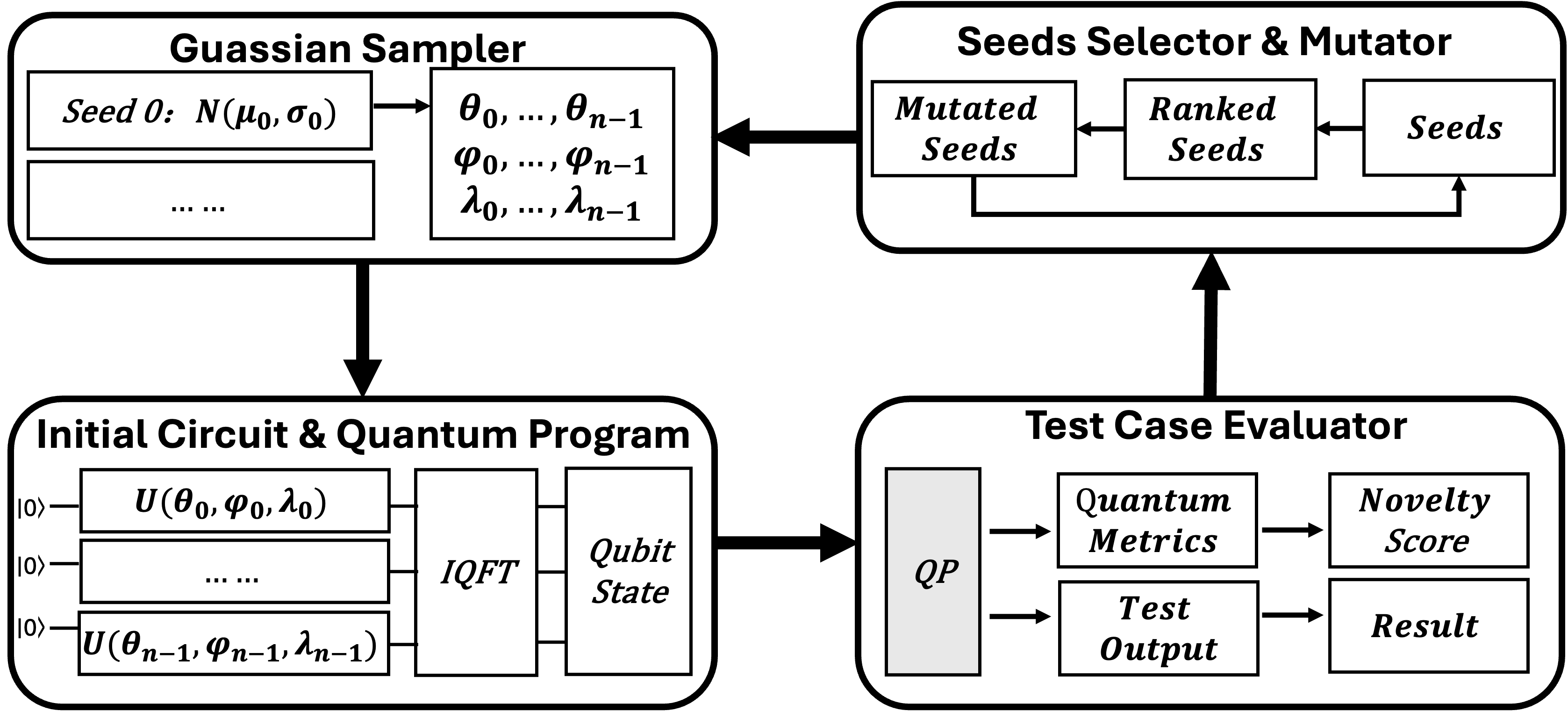}
    \caption{Workflow of \textit{NovaQ}. The module labeled with a shadow (i.e., QP) represents the quantum programs under test.}
    \label{fig:main}
\vspace{-1mm}
\end{figure}

\textit{NovaQ} consists of two main components: a test case generator that produces qubit states based on quantum circuits equipped with parameters sampled from Gaussian distributions and a test case evaluator that quantifies diversity using internal quantum state metrics. The testing process iteratively mutates parameter distributions, generates test circuits, evaluates their behavioral novelty, and retains the most promising distributions for further mutation. The overall workflow is shown in Figure~\ref{fig:main}.

\subsection{Distribution-Based Test Case Generation}

The core of \textit{NovaQ}'s test case generation lies in constructing parameterized initial quantum circuits. Each seed is defined as a pair of real numbers that represent the mean and variance of a normal distribution \( N(\mu, \sigma) \). Samples drawn from these distributions are used to parameterize the \( U \) gates in the initial circuit. Each circuit consists of these parameterized gates followed by a fixed Inverse Quantum Fourier Transform (IQFT) layer to introduce complex interference patterns.

By setting different values for the \( U \) gate parameters \(\theta\), \(\phi\), and \(\lambda\), applying these gates to each qubit initialized in the \(\ket{0}\) state results in qubit states with varied characteristics. The initial circuit takes the all-zero quantum state \( \ket{0}^{\otimes n} \) as input and outputs a qubit state vector. This vector is then used as the input to the quantum program under test. Without the use of diverse initial circuits, the generated test inputs would lack variation, significantly reducing the likelihood of exposing bugs. We formally define a test case as follows:

\begin{definition}
A \textit{test case} refers to the output qubit state vector of the initial quantum circuit—comprising parameterized \( U \) gates and an IQFT layer—when executed on the all-zero input state \( \ket{0}^{\otimes n} \).
\end{definition}

To ensure diversity in test inputs, \textit{NovaQ} initializes multiple seed distributions and applies small mutations to them in each iteration. For each seed, a fixed number of test circuits is generated and evaluated for behavioral diversity and fault-detection potential.

\begin{definition}
A \textit{bug} in a quantum program refers to a violation of the expected input-output correspondence. In a correct quantum program, the output probability distribution should remain consistent for a given input; any deviation from this expected behavior may indicate the presence of a bug.
\end{definition}

\subsection{Diversity-Based Evaluation and Bug Detection}

To effectively identify faults in quantum programs, it is crucial to explore the qubit state input space thoroughly. For each test case, the execution result on the target quantum program is analyzed to extract internal quantum state features—specifically, phase, magnitude, and entanglement—using vector simulations. 
To prevent generated test cases from trapping in specific regions of the input space, we design the \textit{novelty score}, which guides the exploration toward underrepresented behaviors in the input space, thereby enhancing test case diversity.

To evaluate novelty, the continuous space defined by the three metrics is discretized into a finite set of grid cells. Each metric is first normalized to the range $[l_d, u_d]$, and then divided into $N$ equal intervals. Thus, each qubit state is mapped to a unique grid cell defined by its discretized magnitude, phase, and entanglement values. The novelty score of a state is inversely related to the frequency with which its corresponding grid cell has been visited: the less frequently visited, the more novel. This computation is formalized as follows:

\begin{equation}
\label{grid}
\eta_d = \left\lfloor \frac{u_d - l_d}{N} \cdot (\phi_d - l_d) \right\rfloor
\end{equation}

Here, $\eta_d$ is the index of the interval for metric $d$, and $\phi_d$ is the metric value (e.g., magnitude, phase, or entanglement). For each qubit state, we obtain the triplet $(\phi_m, \phi_p, \phi_e)$ representing its metrics, which is then mapped to the discrete grid cell $(\eta_m, \eta_p, \eta_e)$.

\begin{definition}
The \textit{novelty score} of a qubit state is defined as the relative visitation frequency $\frac{\rho}{N^3}$ of its corresponding grid cell, where $\rho$ is the number of previously recorded states mapped to that cell. A lower novelty score indicates that the state lies in a region of the metric space that is rarely explored.
\end{definition}

\subsection{Iterative Selection and Optimization}

The test generation process follows an iterative mutation-selection loop. After generating and evaluating the test circuits from the current seed set, each seed is scored based on the average novelty of its corresponding test cases. The top-performing seeds are selected to form the next generation, guiding the search toward distributions that yield diverse and potentially fault-revealing inputs. Specifically, the mean and variance are sampled within the ranges \([-15, 15]\) and \([0.1, 30]\), respectively. Each mutation perturbs the values of mean and variance as follows:
\[
\mu' = \mu + \Delta \mu, \quad \Delta \mu \in [-0.5, 0.5]
\]
\[
\sigma'^2 = \sigma^2 + \Delta \sigma^2, \quad \Delta \sigma^2 \in [-0.5, 0.5]
\]

This process continues until a user-defined budget (e.g., total number of test cases) is reached.

\section{Evaluation}

We evaluated the effectiveness of \textit{NovaQ} by comparing it with a baseline method in terms of test case diversity and bug detection capability. The evaluation follows the testing methodology described in~\cite{ye2023quratest}, and measures the fault-detection performance of test cases generated by both the baseline and \textit{NovaQ}.

\textbf{RQ1:} Does \textit{NovaQ} generate more diverse test cases compared to the baseline?

To answer this question, we compare the diversity of test cases generated by \textit{NovaQ} with that of the baseline. Diversity is evaluated using a discretized $10 \times 10 \times 10$ grid on three quantum-specific metrics: magnitude, phase, and entanglement, as proposed in~\cite{ye2023quratest}. A higher number of occupied grid cells indicates greater diversity.

We apply both methods to the IQFT generator. While the baseline study in~\cite{ye2023quratest} reports results only for 5-qubit circuits, our evaluation includes circuits with 3, 5, 7, 10, and 12 qubits, for a broader comparison between different circuit sizes.

In \textit{NovaQ}, the initial seed pool contains $n = 10$ randomly initialized seeds. For each iteration, we apply controlled mutations to the three parameters $(\theta, \phi, \lambda)$ of all $U$ gates in the selected seeds. The mutation range for both mean and variance of the parameter distributions is $\pm 0.5$. To avoid premature convergence to local optima, each seed selected for the next generation has a 10\% probability of undergoing random mutation instead of guided mutation.

After generating all $N = 1500$ test cases, we compute the diversity by counting the number of occupied grid cells in the three-dimensional space defined by the three metrics. The results are summarized in Table~\ref{Grid}. Across all qubit sizes, \textit{NovaQ} consistently generates more diverse test cases than the baseline. For qubit numbers of 3, 5, 7, 10, and 12, \textit{NovaQ} achieves improvements of 10.57\%, 13.07\%, 39.40\%, 55.73\%, and 107.37\%, respectively, in test case diversity compared to the baseline. Because the input space expands exponentially with the number of qubits, leading to an increased proportion of certain grid regions within the space. \textcolor{black}{The baseline’s random parameter generation often becomes trapped in high-proportion grids, \textit{NovaQ} leverages a novelty-driven mechanism to explore uncovered regions, thereby attaining substantially higher coverage, especially for larger qubit systems.}


\begin{table}[htbp]
\centering
\caption{Grid Coverage in 1,500 tests of Baseline vs. \textit{NovaQ}}
\begin{scriptsize}
\begin{tabular}{|c||cc|cc|}
\hline
\textbf{Qubit Number} & \multicolumn{2}{c|}{\textbf{Baseline}} & \multicolumn{2}{c|}{\textbf{NovaQ}} \\
\cline{2-5}
 & Grids & Coverage Rate & Grids & Coverage Rate \\
\hline
3  & 634 & 63.4\% & 701 & 70.1\% \\
5  & 574 & 57.4\% & 649 & 64.9\% \\
7  & 434 & 43.4\% & 605 & 60.5\% \\
10 & 192 & 19.2\% & 299 & 29.9\% \\
12 &  95 &  9.5\% & 197 & 19.7\% \\
\hline
\end{tabular}
\label{Grid}
\end{scriptsize}
\end{table}

\textbf{RQ2:} In what way does \textit{NovaQ} outperform the baseline?

\begin{figure}[ht]
    \centering

    \begin{minipage}{0.485\textwidth}
        \includegraphics[width=\textwidth]{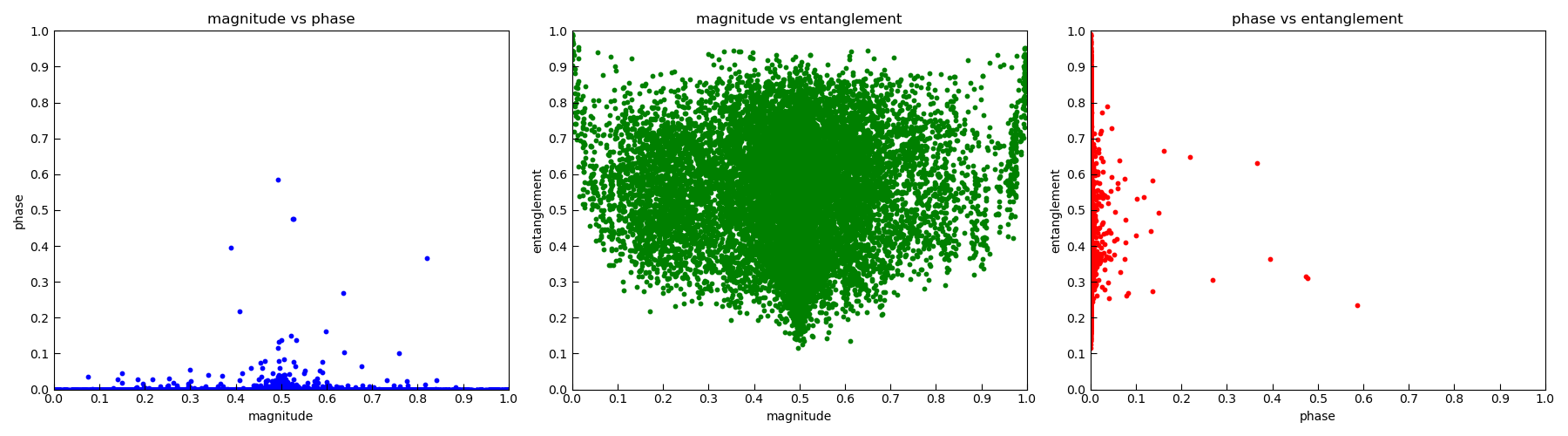}
        \subcaption{Scores of Baseline}
        \label{fig:scores-baseline}
    \end{minipage}

    \par\vspace{1em}  

    \begin{minipage}{0.485\textwidth}
        \includegraphics[width=\textwidth]{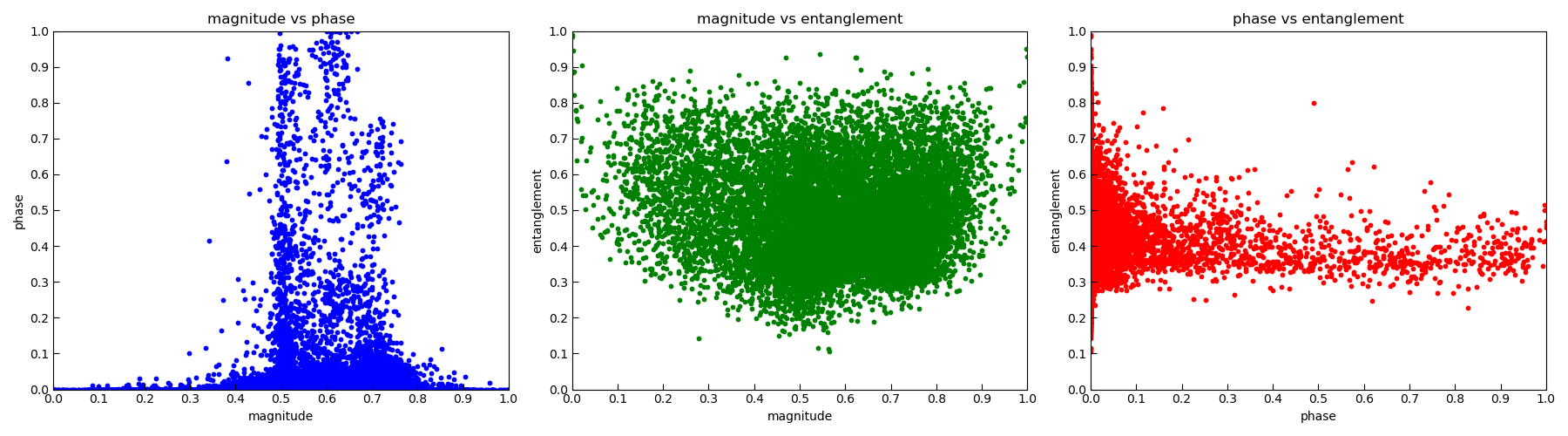}
        \subcaption{Scores of \textit{NovaQ}}
        \label{fig:scores-NovaQ}
    \end{minipage}

    \caption{Diversity difference between Baseline and \textit{NovaQ} of 12-qubit. The blue, green, and red parts represent the results based on the dimensions of (magnitude, phase), (magnitude, entanglement), and (phase, entanglement), respectively.}
    \label{fig:scores}
\end{figure}

To further analyze the results, we use the 12-qubit case from RQ1 as a representative example. In Figure~\ref{fig:scores}, we use three three-view diagrams to show the experimental results, which were originally three-dimensional diagrams. Figures~\ref{fig:scores-baseline} and~\ref{fig:scores-NovaQ} show the distribution of test cases across the three diversity metrics. For magnitude and entanglement scores, both methods yield similar results. However, in the phase dimension, \textit{NovaQ} generates more diverse test cases, leading to higher overall diversity. This demonstrates that \textit{NovaQ}'s novelty-driven mechanism is effective in guiding test generation toward under-explored areas of the state space. \textcolor{black}{Furthermore, the results indicate that \textit{NovaQ}’s effectiveness in \textbf{RQ1}  mainly comes from enhanced grid coverage of phase scores.}


\textbf{RQ3:} Are the test cases generated by \textit{NovaQ} more effective in detecting bugs in quantum programs compared to those generated by the baseline?

To answer this question, we evaluate the fault-detection capability of test cases produced by both \textit{NovaQ} and the baseline. Following the methodology in~\cite{ye2023quratest}, 
we randomly replace certain quantum gates in the common quantum algorithms with arbitrary quantum gates and filter out buggy programs that affect the original functionality according to \textit{Definition 2} as the benchmark. We then measure the detection rate for each test suite generated by \textit{NovaQ} and the baseline.
A higher detection rate indicates stronger fault sensitivity and better testing effectiveness.

\begin{table}[htbp]
\vspace{-1mm}
\centering
\caption{Number of bugs found in 1,500 tests of Baseline vs. \textit{NovaQ}}
\begin{scriptsize}
\begin{tabular}{|c||cc|cc|}
\hline
\textbf{Program} & \multicolumn{2}{c|}{\textbf{Baseline}} & \multicolumn{2}{c|}{\textbf{NovaQ}} \\
\cline{2-5}
 & Bugs Found & Accuracy & Bugs Found & Accuracy \\
\hline
Grover-03  & 1270 & 84.7\% & 1367 & 91.1\% \\
Grover-05  & 1258 & 83.9\% & 1405 & 93.7\% \\
Grover-07  & 1249 & 83.3\% & 1391 & 92.7\% \\
Grover-10  & 1280 & 85.3\% & 1426 & 95.1\% \\
Grover-12  & 1280 & 85.5\% & 1412 & 92.1\% \\
PE-05  & 1251 & 83.4\% & 1387 & 92.5\% \\
QFT-05  & 1277 & 85.1\% & 1341 & 89.4\% \\
\hline
\end{tabular}
\label{resultgrover}
\end{scriptsize}
\vspace{-1mm}
\end{table}

We conducted experiments using a faulty implementation of Grover's algorithm as a benchmark. In each run, we generate 1,500 test cases and count the number of faulty programs correctly identified. The results are summarized in Table~\ref{resultgrover}. Across all tested circuit sizes, \textit{NovaQ} detects significantly more bugs than the baseline. For example, in the 12-qubit setting, \textit{NovaQ} detects 1412 faults (92.1\% accuracy), compared to 1280 faults (85.5\% accuracy) by the baseline. \textcolor{black}{This is because a quantum program may involve many possible execution branches, and faults can occur in any of the branches. By increasing the diversity of test cases, a wider range of branch combinations can be exercised, thereby improving the chances of exposing hidden bugs.}

\vspace{1mm}
\textbf{RQ4:} How do the growth rates of Grid compare under the two methods?

\begin{figure}[htbp]
    \centering
    \includegraphics[width=0.95\linewidth]{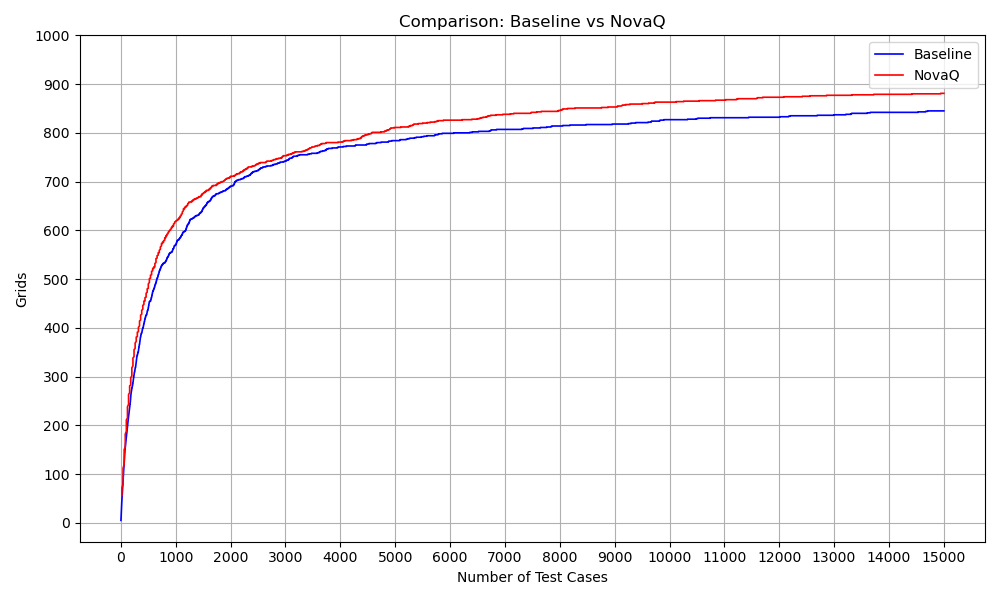}
    \caption{15,000 Test Case Results of 3-qubit: Baseline vs \textit{NovaQ}}
    \label{fig:15000}
\end{figure}

To answer this question, we select the results of 15,000 test cases generated when the number of qubits is 3 for baseline and \textit{NovaQ}, and plot a graph showing the relationship between the number of grids and the number of test cases. As Figure~\ref{fig:15000} shows, \textit{NovaQ} has a higher growth rate than the baseline in generating more types of test cases. Furthermore, when generating the same number of test cases, \textit{NovaQ} consistently generates more diverse test cases than the baseline.

Figure~\ref{fig:15000} also shows that when generating a particularly large number of test cases, although it is difficult for the new test cases generated by baseline and \textit{NovaQ} to become more diverse, the upper limit of diversity of the test cases generated by \textit{NovaQ} is higher than that of the baseline. This indicates that the mutation parameter method in \textit{NovaQ} can generate types of test cases that cannot be generated by the pure random method in the baseline.
\section{Related Work}

Quantum software testing faces fundamental challenges due to the non-deterministic nature of quantum programs. Several studies~\cite{garcia2023quantum, paltenghi2024survey, leite2025testing} have surveyed the current landscape of testing approaches and challenges. To support systematic investigation, benchmark datasets have been developed. Bugs4Q~\cite{zhao2021bugs4q,zhao2023bugs4q} collects 36 validated bugs from Qiskit and extends to 42 from GitHub and community platforms, while QBugs~\cite{campos2021qbugs} provides a framework for organizing and reproducing quantum software bugs in a controlled setting. Building on these resources, various testing techniques have been explored, including search-based testing~\cite{10.1145/3510454.3516839}, combinatorial testing~\cite{9724888}, metamorphic testing~\cite{abreu2022metamorphic}, concolic testing~\cite{xia2025quantum}, mutation testing~\cite{fortunato2022qmutpy}, and black-box testing~\cite{long2025black,long2024equivalence}.

Beyond these methods, test case generation, a crucial step in classical software testing, has also been adapted for quantum programs. Coverage-based approaches have been proposed to generate effective test cases~\cite{wang2021quito, 10.1145/3510454.3516839, 9724888}, and QuraTest~\cite{ye2023quratest} leverages quantum properties to guide generation, though existing methods have not fully exploited such properties to enhance diversity. NovaQ addresses this limitation by adopting a diversity-guided strategy to select more informative test cases. Complementary to test generation, several quantum-specific coverage criteria~\cite{ali2021assessing,fortunato2024gate, long2024testing,shao2024coverage} have been introduced to support systematic evaluation, and tools and frameworks~\cite{fortunato2022mutation, pontolillo2024delta,long2024testing,li2020projection} have been developed to facilitate test execution and assertion checking, together forming a growing ecosystem of quantum testing support.
\section{Conclusion}

This paper presents \textit{NovaQ}, a testing framework tailored for quantum programs. Unlike existing approaches that rely on random parameter generation, \textit{NovaQ} employs a mutation-based strategy that perturbs the mean and variance of parameter distributions during test case generation to improve diversity. 
%
Experimental results show that across various qubit numbers, \textit{NovaQ} outperforms the baseline method in terms of test case diversity and fault detection. Moreover, \textit{NovaQ} achieves more effective exploration in the phase property.

Future work will explore applying \textit{NovaQ} to additional test case generators beyond the UCNOT and IQFT methods, and evaluate its effectiveness across a broader range of quantum programs to further validate the impact of test case diversity on fault detection.


\bibliographystyle{IEEEtran}
\bibliography{ref}

\begin{thebibliography}{10}
\providecommand{\url}[1]{#1}
\csname url@samestyle\endcsname
\providecommand{\newblock}{\relax}
\providecommand{\bibinfo}[2]{#2}
\providecommand{\BIBentrySTDinterwordspacing}{\spaceskip=0pt\relax}
\providecommand{\BIBentryALTinterwordstretchfactor}{4}
\providecommand{\BIBentryALTinterwordspacing}{\spaceskip=\fontdimen2\font plus
\BIBentryALTinterwordstretchfactor\fontdimen3\font minus \fontdimen4\font\relax}
\providecommand{\BIBforeignlanguage}[2]{{%
\expandafter\ifx\csname l@#1\endcsname\relax
\typeout{** WARNING: IEEEtran.bst: No hyphenation pattern has been}%
\typeout{** loaded for the language `#1'. Using the pattern for}%
\typeout{** the default language instead.}%
\else
\language=\csname l@#1\endcsname
\fi
#2}}
\providecommand{\BIBdecl}{\relax}
\BIBdecl

\bibitem{kitaev2002classical}
A.~Y. Kitaev, A.~Shen, M.~N. Vyalyi, and M.~N. Vyalyi, \emph{Classical and quantum computation}.\hskip 1em plus 0.5em minus 0.4em\relax American Mathematical Soc., 2002, no.~47.

\bibitem{gisin2002quantum}
N.~Gisin, G.~Ribordy, W.~Tittel, and H.~Zbinden, ``Quantum cryptography,'' \emph{Reviews of modern physics}, vol.~74, no.~1, p. 145, 2002.

\bibitem{abbas2024challenges}
A.~Abbas, A.~Ambainis, B.~Augustino, A.~B{\"a}rtschi, H.~Buhrman, C.~Coffrin, G.~Cortiana, V.~Dunjko, D.~J. Egger, B.~G. Elmegreen \emph{et~al.}, ``Challenges and opportunities in quantum optimization,'' \emph{Nature Reviews Physics}, pp. 1--18, 2024.

\bibitem{georgescu2014quantum}
I.~M. Georgescu, S.~Ashhab, and F.~Nori, ``Quantum simulation,'' \emph{Reviews of Modern Physics}, vol.~86, no.~1, pp. 153--185, 2014.

\bibitem{piani2016robustness}
M.~Piani, M.~Cianciaruso, T.~R. Bromley, C.~Napoli, N.~Johnston, and G.~Adesso, ``Robustness of asymmetry and coherence of quantum states,'' \emph{Physical Review A}, vol.~93, no.~4, p. 042107, 2016.

\bibitem{harrow2003robustness}
A.~W. Harrow and M.~A. Nielsen, ``Robustness of quantum gates in the presence of noise,'' \emph{Physical Review A}, vol.~68, no.~1, p. 012308, 2003.

\bibitem{jin2023scaffml}
T.~Jin and J.~Zhao, ``Scaffml: A quantum behavioral interface specification language for scaffold,'' in \emph{2023 IEEE International Conference on Quantum Software (QSW)}.\hskip 1em plus 0.5em minus 0.4em\relax IEEE, 2023, pp. 128--137.

\bibitem{streltsov2012quantum}
A.~Streltsov, H.~Kampermann, and D.~Bru{\ss}, ``Quantum cost for sending entanglement,'' \emph{Physical review letters}, vol. 108, no.~25, p. 250501, 2012.

\bibitem{gottesman2002introduction}
D.~Gottesman, ``An introduction to quantum error correction,'' in \emph{Proceedings of Symposia in Applied Mathematics}, vol.~58, 2002, pp. 221--236.

\bibitem{leite2025testing}
N.~C. Leite~Ramalho, H.~Amario~de Souza, and M.~Lordello~Chaim, ``Testing and debugging quantum programs: The road to 2030,'' \emph{ACM Transactions on Software Engineering and Methodology}, vol.~34, no.~5, pp. 1--46, 2025.

\bibitem{miranskyy2019testing}
A.~Miranskyy and L.~Zhang, ``On testing quantum programs,'' in \emph{2019 IEEE/ACM 41st International Conference on Software Engineering: New Ideas and Emerging Results (ICSE-NIER)}.\hskip 1em plus 0.5em minus 0.4em\relax IEEE, 2019, pp. 57--60.

\bibitem{paltenghi2024survey}
M.~Paltenghi and M.~Pradel, ``A survey on testing and analysis of quantum software,'' \emph{arXiv preprint arXiv:2410.00650}, 2024.

\bibitem{ye2023quratest}
J.~Ye, S.~Xia, F.~Zhang, P.~Arcaini, L.~Ma, J.~Zhao, and F.~Ishikawa, ``Quratest: Integrating quantum specific features in quantum program testing,'' in \emph{2023 38th IEEE/ACM International Conference on Automated Software Engineering (ASE)}.\hskip 1em plus 0.5em minus 0.4em\relax IEEE, 2023, pp. 1149--1161.

\bibitem{zhang2016gaussian}
X.~Zhang, ``Gaussian distribution,'' in \emph{Encyclopedia of machine learning and data mining}.\hskip 1em plus 0.5em minus 0.4em\relax Springer, 2016, pp. 1--5.

\bibitem{weinstein2001implementation}
Y.~S. Weinstein, M.~Pravia, E.~Fortunato, S.~Lloyd, and D.~G. Cory, ``Implementation of the quantum fourier transform,'' \emph{Physical review letters}, vol.~86, no.~9, p. 1889, 2001.

\bibitem{garcia2023quantum}
A.~Garc{\'\i}a de~la Barrera, I.~Garc{\'\i}a-Rodr{\'\i}guez~de Guzm{\'a}n, M.~Polo, and M.~Piattini, ``Quantum software testing: State of the art,'' \emph{Journal of Software: Evolution and Process}, vol.~35, no.~4, p. e2419, 2023.

\bibitem{zhao2021bugs4q}
P.~Zhao, J.~Zhao, Z.~Miao, and S.~Lan, ``Bugs4{Q}: A benchmark of real bugs for quantum programs,'' in \emph{2021 36th IEEE/ACM International Conference on Automated Software Engineering (ASE)}.\hskip 1em plus 0.5em minus 0.4em\relax IEEE, 2021, pp. 1373--1376.

\bibitem{zhao2023bugs4q}
P.~Zhao, Z.~Miao, S.~Lan, and J.~Zhao, ``Bugs4{Q}: A benchmark of existing bugs to enable controlled testing and debugging studies for quantum programs,'' \emph{Journal of Systems and Software}, vol. 205, p. 111805, 2023.

\bibitem{campos2021qbugs}
J.~Campos and A.~Souto, ``Qbugs: A collection of reproducible bugs in quantum algorithms and a supporting infrastructure to enable controlled quantum software testing and debugging experiments,'' in \emph{2021 IEEE/ACM 2nd International Workshop on Quantum Software Engineering (Q-SE)}.\hskip 1em plus 0.5em minus 0.4em\relax IEEE, 2021, pp. 28--32.

\bibitem{10.1145/3510454.3516839}
\BIBentryALTinterwordspacing
X.~Wang, P.~Arcaini, T.~Yue, and S.~Ali, ``Qusbt: search-based testing of quantum programs,'' in \emph{Proceedings of the ACM/IEEE 44th International Conference on Software Engineering: Companion Proceedings}, ser. ICSE '22.\hskip 1em plus 0.5em minus 0.4em\relax New York, NY, USA: Association for Computing Machinery, 2022, p. 173–177. [Online]. Available: \url{https://doi.org/10.1145/3510454.3516839}
\BIBentrySTDinterwordspacing

\bibitem{9724888}
------, ``Application of combinatorial testing to quantum programs,'' in \emph{2021 IEEE 21st International Conference on Software Quality, Reliability and Security (QRS)}, 2021, pp. 179--188.

\bibitem{abreu2022metamorphic}
R.~Abreu, J.~P. Fernandes, L.~Llana, and G.~Tavares, ``Metamorphic testing of oracle quantum programs,'' in \emph{Proceedings of the 3rd International Workshop on Quantum Software Engineering}, 2022, pp. 16--23.

\bibitem{xia2025quantum}
S.~Xia, J.~Zhao, F.~Zhang, and X.~Guo, ``Quantum concolic testing,'' \emph{Proceedings of the ACM on Software Engineering}, vol.~2, no. ISSTA, pp. 1146--1166, 2025.

\bibitem{fortunato2022qmutpy}
D.~Fortunato, J.~Campos, and R.~Abreu, ``Qmutpy: A mutation testing tool for quantum algorithms and applications in qiskit,'' in \emph{Proceedings of the 31st ACM SIGSOFT International Symposium on Software Testing and Analysis}, 2022, pp. 797--800.

\bibitem{long2025black}
P.~Long and J.~Zhao, ``A black-box testing framework for oracle quantum programs,'' \emph{arXiv preprint arXiv:2505.07243}, 2025.

\bibitem{long2024equivalence}
------, ``Equivalence, identity, and unitarity checking in black-box testing of quantum programs,'' \emph{Journal of Systems and Software}, vol. 211, p. 112000, 2024.

\bibitem{wang2021quito}
X.~Wang, P.~Arcaini, T.~Yue, and S.~Ali, ``Quito: a coverage-guided test generator for quantum programs,'' in \emph{2021 36th IEEE/ACM International Conference on Automated Software Engineering (ASE)}.\hskip 1em plus 0.5em minus 0.4em\relax IEEE, 2021, pp. 1237--1241.

\bibitem{ali2021assessing}
S.~Ali, P.~Arcaini, X.~Wang, and T.~Yue, ``Assessing the effectiveness of input and output coverage criteria for testing quantum programs,'' in \emph{2021 14th IEEE Conference on Software Testing, Verification and Validation (ICST)}.\hskip 1em plus 0.5em minus 0.4em\relax IEEE, 2021, pp. 13--23.

\bibitem{fortunato2024gate}
D.~Fortunato, J.~Campos, and R.~Abreu, ``Gate branch coverage: A metric for quantum software testing,'' in \emph{Proceedings of the 1st ACM International Workshop on Quantum Software Engineering: The Next Evolution}, 2024, pp. 15--18.

\bibitem{long2024testing}
P.~Long and J.~Zhao, ``Testing multi-subroutine quantum programs: From unit testing to integration testing,'' \emph{ACM Transactions on Software Engineering and Methodology}, vol.~33, no.~6, pp. 1--61, 2024.

\bibitem{shao2024coverage}
M.~Shao and J.~Zhao, ``A coverage-guided testing framework for quantum neural networks,'' \emph{arXiv preprint arXiv:2411.02450}, 2024.

\bibitem{fortunato2022mutation}
D.~Fortunato, J.~Campos, and R.~Abreu, ``Mutation testing of quantum programs: A case study with qiskit,'' \emph{IEEE Transactions on Quantum Engineering}, vol.~3, pp. 1--17, 2022.

\bibitem{pontolillo2024delta}
G.~J. Pontolillo and M.~R. Mousavi, ``Delta debugging for property-based regression testing of quantum programs,'' in \emph{Proceedings of the 5th ACM/IEEE International Workshop on Quantum Software Engineering}, 2024, pp. 1--8.

\bibitem{li2020projection}
G.~Li, L.~Zhou, N.~Yu, Y.~Ding, M.~Ying, and Y.~Xie, ``Projection-based runtime assertions for testing and debugging quantum programs,'' \emph{Proceedings of the ACM on Programming Languages}, vol.~4, no. OOPSLA, pp. 1--29, 2020.

\end{thebibliography}

\end{document}